# Social Norm Design for Information Exchange Systems with Limited Observations

Jie Xu and Mihaela van der Schaar[1]

***Abstract*— Everyday people turn to the web to exchange services, data and ideas on websites such as BitTorrent, Yahoo Answers, Yelp, Amazon Mechanical Turk and more. These information exchange systems differ in many ways, but all share a common vulnerability to selfish behavior and free-riding. In this paper, we build incentives schemes based on social norms. Social norms prescribe a social strategy for the users in the system to follow and deploy reputation schemes to reward or penalize users depending on whether they follow or deviate from the prescribed strategy when selecting actions. Because users in these systems often have only limited capability to observe the global system information, e.g. the reputation distribution of the users participating in the system, their beliefs about the reputation distribution are heterogeneous and biased. Such belief heterogeneity causes a positive fraction of users to not follow the social strategy. In such practical scenarios, the standard equilibrium analysis deployed in the economics literature is no longer directly applicable and hence, the system design needs to consider these differences. To investigate how the system designs need to change when the participating users have only limited observations, we focus on a simple social norm with binary reputation labels but allow adjusting the punishment severity through randomization. First, we model the belief heterogeneity using a suitable Bayesian belief function. Next, we formalize the users' optimal decision problems and derive in which scenarios they follow the prescribed social strategy. With this result, we then study the system dynamics and formally define equilibrium in the sense that the system is stable when users strategically optimize their decisions. By rigorously studying two specific cases where users' belief distribution is constant or is linearly influenced by the true reputation distribution, we prove that the optimal reputation update rule is to choose the mildest possible punishment. This result is further confirmed for higher order beliefs in simulations. It is also shown that more observations do not necessarily lead to a higher efficiency. In conclusion, our proposed design framework enables the development of optimal social norms for various deployment scenarios with limited observations.**

## I. Introduction

As the web has evolved, it has become increasingly social. People turn to the web to exchange ideas, data and services, as evidenced by the popularity of sites like Wikipedia, BitTorrent, Yahoo Answers, Yelp, and online labor markets like Amazon Mechanical Turk (AMT). While these systems, which we refer to as *information exchange systems*, differ in many ways, they share a common vulnerability to selfish behavior and free-riding. For example, a worker on AMT may attempt to complete jobs with as little effort as possible while still being paid; a user in a peer-to-peer system may wish to download files without using bandwidth to upload files for others. In order for these sites to thrive, participants must be properly motivated to contribute.

Distributed optimization techniques have been applied extensively in engineering to enable the efficient usage of resources by obedient or cooperative users. Only in recent years have engineers started to investigate incentive issues in systems formed by self-interested users. Many of the existing mechanisms to combat free-riding problems rely on game-theoretic approaches and can be classified as either *pricing* mechanisms or *reciprocity* mechanisms. Pricing mechanisms are appropriate in some settings, but do not make sense for applications like Yahoo Answers, Wikipedia, or Yelp, where much of the appeal is that the information is free.

[1] The authors are with Department of Electrical Engineering, University of California, Los Angeles (UCLA). (Email: jiexu@ucla.edu, mihaela@ee.ucla.edu.)

Under a reciprocity mechanism, a user is rewarded or punished based on its behavior in the system. Rewards and punishment are typically determined according to a differential service scheme [1], which might require, for example, that a user who contributed heavily to the system in the past should receive more resources than a user who contributed less [2]. This preferential treatment provides an incentive for users to cooperate, and can be implemented using either virtual currency or reputation. Under a virtual currency mechanism, users are incentivized to contribute through a system of rewards based on virtual currency [3][4][5][6]. However, prior work shows that even optimal designs based on virtual currency cannot achieve optimal performance [7]. To measure good behavior, reciprocity mechanisms frequently associate a rating or reputation score with each other in the system. Depending on how a user's rating is generated, reciprocity-based protocols can be classified as direct reciprocity mechanisms [8], or indirect reciprocity mechanisms [9]. Direct reciprocity implies that the interaction between two users is influenced only by their history of interactions with each other, and not by their interactions with other users. Though easy to implement, direct reciprocity requires frequent interactions between two users in order to establish accurate mutual ratings. This is restrictive in systems characterized by high churn, asymmetry of interests, or infrequent interactions between any pair of users, such as most peer production systems, online labor markets, and review sites.

Protocols that are based on indirect reciprocity typically assign to each user a global reputation [10] based on its past interactions with all other users in the system. A differential service scheme recommends actions (e.g., "share a file with this user" or "do not share a file with this user") based only on the reputations of users, and not on their entire history of interactions. Much of the existing work on reputation mechanisms is concerned with practical implementation details. Some focuses on effective information gathering techniques which differ in how the global reputation is calculated and propagated (e.g., through efficient information aggregation [11] or secure user identification [12]). Empirical studies have examined the impact of reputation on a seller's prices and sales [13][14][15], motivation for participating in reputation-based mechanisms [16], among other things. There has also been some work analytically exploring the use of reputation mechanisms to combat moral hazard in a repeated games setting [17][18][19], including some that does not require the presence of a trusted centralized system [20]. This work typically considers one (or a few) long-lived seller with many short-lived buyers, which is not appropriate for information exchange systems where there are many interacting users playing the role of buyer or seller or both, contributing and seeking information.

To rigorously capture the impact of various strategy and protocol design choices on information exchange systems, the authors of [21] propose a framework using *social norms* which were originally designed to sustain cooperation in a community with a large population of individuals participating in anonymous random matching games [22][23][24]. In an incentive scheme based on a social norm, each individual is assigned a dynamic label indicating its reputation or status based on past behavior, and individuals with different labels are treated differently by others in the system. Hence, a social norm can be adopted easily in social communities with an infrastructure that collects, processes, and delivers information about individuals' behavior.

We build incentive schemes for information exchange systems based on social norms. In information exchange systems, users often have imperfect knowledge of the global information, in particular, the reputation distribution of the participating users. For example, users observe the reputations of a limited number of other users on the website and form (probably biased) beliefs of the reputation distribution. Moreover, users' beliefs are heterogeneous since the observations of various users are different. Such belief heterogeneity causes a positive fraction of users to not follow the social strategy. In contrast, standard equilibrium analysis [21][23] requires that all users follow the social strategy and is conducted under two assumptions: (1) users have homogenous and accurate knowledge about the reputation distribution; and (2) users believe that all users obey the social strategy. These assumptions [21][23] hold only if users have unlimited observations and hence, they have perfect knowledge about the reputation distribution of the participating users. However, they do not hold in many practical systems where users only have limited observations and the system dynamics does not evolve to an equilibrium where all users follow the social strategy. Instead users have heterogeneous beliefs about the reputation distribution and they tend to trust other users with high reputations and distrust those with low reputations. Therefore, users' limited observation capability leads to a different system design. The main contributions of this paper are summarized as follows:

- We propose a simple class of social norms with binary reputations but allow adjusting the punishment severity through randomization. This class is simple and easy to implement while has also been shown to be close to the optimal strategy for the unlimited observations case in [21]. (Note that this strategy includes the contagion strategy [23] as a special case.) Similar randomization approach is also used in [22].
- We model the users' heterogeneous beliefs of the reputation distribution due to limited observations using a Bayesian belief model, which captures the feature that the observation depends on the current true reputation distribution and that more observations lead to more accurate information about the reputation distribution.
- We prove that users follow the social strategy only if their beliefs about the reputation distribution are above certain thresholds, i.e., they need to have sufficient "trust" in the society. Using this result, we can show that, in most interesting scenarios, the optimal design is to use the mildest possible punishment, thereby leading to a different social norm design than in the unlimited observations cases.

The rest of this paper is organized as follows. Section II describes the basic setup, the social norms and builds the belief model. Section III investigates users' decision problem. System dynamics and the equilibrium are then studied. In Section IV, the impact of punishment on the equilibrium performance is investigated. The optimal design is derived for two specific Bayesian belief functions. Simulations are conducted in Section V followed by conclusions in Section VI.

## II. System Model

### A. Setup

We consider an information exchange system where users request and provide information or resources. We utilize the widely-used continuum model (mass 1), implicitly assuming that the user population is large and static. The system is modeled as a discrete-time system where time is divided into periods. When a requester generates a task, it is posted on the website and a provider is assigned to solve the task. We assume that there is no price associated with the task (as in Yelp, Yahoo Answers and etc.), the provider is the only strategic part that needs to decide whether or not to exert effort to solve the task. Upon accepting, the provider incurs a cost $c$ to fulfill the task while the requester receives a benefit $b$. We assume $b > c > 0$ to make providing the service socially valuable and denote $\gamma = b / c$ as the benefit-to-cost ratio. This is a simple gift-giving game (see Fig. 1) in which the dominant strategy for the provider is not to provide service. Incentives can be provided if the provider is long-lived in the system and will also become a requester in the future. We assume that users discount the future utility by a constant rate $\beta \in (0,1)$. For the accurate modeling for the real systems, we assume that in each period, each user requests a task to be solved and another user is randomly assigned to solve this task. This random matching model is common in the economics literature [22][23][24]. Nevertheless, the analysis could also apply to the scenarios where a fraction $\lambda \in [0,1]$ of the population generates tasks in each period. The parameter $\lambda$ only indicates the request arrival rate of the system but does not change the result. For the considered case $\lambda = 1$, each user is a requester as well as a provider who is assigned to another user's requested task.

### B. Punishment adjustable social norm

A social norm $\kappa$, which is designed by the protocol designer, is composed of a social strategy $\sigma$, a reputation update rule $\tau$, and a reputation set $\Theta$. Each user is tagged with a reputation $\theta$ representing its social status. We consider only two available reputation labels for the users $\Theta = \{0,1\}$ with $\theta = 1$ indicating a good status and $\theta = 0$ indicating a bad status. Denote the social strategy by the mapping $\sigma : \Theta \to \mathcal{A}$, where $\Theta$ is the reputation set of the requester and $\mathcal{A} = \{0,1\}$ stands for the action set of the provider. The action $a = 1$ represents the case where the provider offers the service while when $a = 0$ it does not provide service. Simply, the social strategy is $\sigma(1) = 1, \sigma(0) = 0$. The social strategy favors good users in such a way that the providers are suggested to provide service only to good requesters but not to provide service to bad requesters. This strategy has the similar merit as the well-known Tit-for-Tat (TfT) strategy in rewarding for cooperative behaviors and punishing for non-cooperative behaviors. However, TfT strategy requires direct reciprocity between interacting users while the social strategy that we use is applicable in systems where users have infrequent interactions and with indirect reciprocity.

The social norm also imposes a reputation update rule based on the action that the provider takes. Intuitively, users who follow the social strategy should receive good reputations and those who do not should receive bad reputations. Denote the reputation update rule by the mapping $\tau : \Theta \times \Theta' \times \mathcal{A} \to [0,1]$, where $\theta \in \Theta$ is the provider's reputation, $\theta' \in \Theta'$ is the requester's reputation and $[0,1]$ indicates the probability that the provider has a good reputation in the next period. The update rule that we use is: $\tau(\theta, 0, a) = \theta, \forall a$ and

$$\tau(1,1,a) = \begin{cases} 1, & \text{if } a = \sigma(1) \\ 0, & \text{if } a \neq \sigma(1) \end{cases} \qquad \tau(0,1,a) = \begin{cases} \alpha, & \text{if } a = \sigma(1) \\ 0, & \text{if } a \neq \sigma(1) \end{cases} \tag{1}$$

Essentially, if the provider deviates from the prescribed social strategy when meeting a good requester[2], its reputation drops to 0; if the bad provider follows the prescribed social strategy, it restores a good reputation with probability $\alpha \in [0,1]$. Hence, for a user to receive service when it becomes a requester in the future, it needs to follow the social strategy as a provider in the current period. The parameter $\alpha$ adjusts the severity of punishment of the social norm which needs to be designed by the system designer. Such randomization can be easily implemented by the central entity that maintains and processes users' reputations. A similar randomization approach is also used in [22] to adjust the punishment severity. For $\alpha = 1$, the punishment is the mildest, allowing the bad provider to repair its reputation after one-time cooperation with a good requester; for $\alpha = 0$, the punishment is the harshest, preventing the bad provider from having a good reputation again in the future no matter how it behaves; for $\alpha \in (0,1)$, the expected time periods for which the user remains in the bad reputation is at least $1/\alpha$. Even though we focus on a system using only binary reputation labels, randomization affects the punishment severity similarly as a system using multiple (more than 2) reputation labels. We portray the aforementioned reputation update rule in Fig. 2.

[2] We will use in the remainder of this paper the term "good/bad" users to refer to the users with good/bad reputations.

### C. *Belief heterogeneity*

In this subsection, we model the users' belief heterogeneity due to their limited observations. Because users are far-sighted, their decisions depend on how they evaluate the status of the society, i.e., the reputation distribution of the system. Since we are considering a binary reputation system, the reputation distribution can be fully described by the fraction of users with good reputations, which we define as the *social reputation* $\rho_s$ and use in the rest of this paper.

In each period, each user observes the system (e.g. observes the reputations of a number of other users) and form a belief of the social reputation. The beliefs are different for different users. Also, note that a user's belief also varies across time because it makes different observations in each period. For a given social reputation $\rho_s$, we model users' beliefs of the social reputation as a probability measure on the support $[0,1]$. Specifically, conditional on the true social reputation, users believe with probability $f(\rho \mid \rho_s)$ that the social reputation is $\rho$. We introduce the observation granularity $M$ to describe how much users are able to observe the system. The interpretation of such granularity can be the number of other users' reputations that the strategic user is able to observe by sampling the website or the number of past interactions that it is able to memorize. The larger $M$ is, the more accurate beliefs that the users have about the social reputation should be. In the following, we describe a belief function that satisfies this property (similar belief function is used in [25] to model users' posterior beliefs after observations):

$$f(\rho \mid \rho_s) = \sum_{m=0}^{M} \binom{M}{m} \rho_s^{\,m} (1-\rho_s)^{M-m} f_{m,M}(\rho), \tag{2}$$

with

$$f_{m,M}(\rho)=\frac{\Gamma(M+2)}{\Gamma(m+1)\Gamma(M-m+1)}\rho^{m}(1-\rho)^{M-m} \quad (3)$$

and where $\Gamma(\cdot)$ denotes the gamma function. Essentially $f_{m,M-m}(\rho)$ is the beta distribution $B(m+1,M-m+1)$. In Bayesian statistics, the beta distribution can be seen as the posterior probability of the parameter $\rho$ of a binomial distribution after observing $m$ successes and $M-m$ failures. Hence, $f_{m,M-m}(\rho)$ can be interpreted as the user's belief of the social reputation after observing $m$ good users and $M-m$ bad users. Furthermore, suppose that the matching process is uniformly random, the number of observations of good users also follows a binomial random distribution with parameter $\rho_s$. With the beta distribution, users' belief distribution is continuous and parameterized by $M$ and $\rho_s$. Let us discuss the extreme cases:

- $M=0$. $f_{m,M}(\rho)=f_{0,0}(\rho)$ is constant and hence, $f(\rho)$ is constant. It implies that users' beliefs of the social reputation are uniformly random.
- $M\to\infty$. In this case, $f(\rho)\to I(\rho-\rho_s)$ where $I(\cdot)$ is the indicate function. It implies that as the observation granularity becomes infinite, users have perfect knowledge of the social reputation. This result is obtained in the following proposition.

**Proposition 1**: For a given $\rho_s$, $\forall\delta$, $\forall\epsilon$, there exists $M(\delta)$ large such that $\forall M\geq M(\delta)$,

$$\int_{\rho=\rho_s-\delta}^{\rho_s+\delta} f(\rho\,|\,\rho_s)\,d\rho\geq 1-\epsilon \quad (4)$$

**Proof**: Omitted due to space limitation. The proof can be found in [27]. ■

## III. System Dynamics and Equilibrium

In this section, we discuss the system dynamics and formally define the Bayesian-Nash equilibrium. For this, we first need to formalize the provider's decision problem and characterize the social reputation distribution that arises in the steady-state in our model.

### *A. User's decision problem*

We begin by investigating a typical provider's decision problem assuming that it has a belief $\rho$. The provider's decision will be based on its own reputation $\theta$, the requester's reputation $\theta'$ and its belief $\rho$ toward the social reputation. The provider chooses an action $a(\theta,\theta'\,|\,\rho)\in\mathcal{A}$ to maximize its total expected discounted utility. Depending on which action the provider takes, the reputation transition follows the reputation update rule. The provider will follow the social strategy if the long-run payoff is larger than the payoff by deviating and will deviate otherwise. For the social strategy $\sigma$, the stage payoff is $\pi(\sigma \mid \theta, \theta', \rho)$ when the provider chooses an action determined by $\sigma$ and holds the belief $\rho$ and the long-run payoff is given by

$$\begin{aligned} &V(\sigma \mid \theta, \theta', \rho) = \pi(\sigma \mid \theta, \theta', \rho) \\ &+\beta\big(\rho V(\sigma \mid \tau(\theta, \theta', \sigma), 1, \rho) + (1-\rho) V(\sigma \mid \tau(\theta, \theta', \sigma), 0, \rho)\big) \end{aligned} \tag{5}$$

As in [26], we assume that the provider believes that bad users play defect and good users follow the social strategy when calculating its payoff. For example, if the provider has a belief $\rho = 0$, it believes that no other user will provide service to itself when it requests service in the future even if it has a good reputation. The next proposition shows that the provider needs to have sufficient "trust" in the society in order for it to be willing to follow the social strategy.

**Proposition 2**: The optimal action $a^*(\theta, \theta' \mid \rho)$ for the provider with belief $\rho$ to follow the prescribed social strategy has a threshold property, i.e.,

$$a^*(1,1 \mid \rho) = \begin{cases} 1 = \sigma(1), & \text{if } \rho \geq \rho_G \\ 0, & \text{if } \rho < \rho_G \end{cases}; \quad a^*(0,1 \mid \rho) = \begin{cases} 1 = \sigma(1), & \text{if } \rho \geq \rho_B \\ 0, & \text{if } \rho < \rho_B \end{cases} \tag{6}$$

and $a^*(\theta, \theta' \mid \rho) = \sigma(\theta')$ for all other cases where $\rho_G, \rho_B$ are threshold beliefs determined by the system parameters $\beta, b, c, \alpha$. Moreover, $\rho_G \leq \rho_B$ and equality holds only if $a = 1$.

**Proof:** (1) First we consider the decision problem when the provider has a good reputation. Obviously, if the requester's reputation is bad, it is optimal that the provider follows the social strategy and plays defect (not provide). If the requester's reputation is good, the provider may have incentives to deviate from the social strategy due to the instant cost. At the decision point, the provider meets a good requester, the expected stage payoff by following the social strategy is $\pi(\sigma) = \rho b - c$ and the stage payoff by deviating is $\pi(\sigma') = \rho b$. Deviation causes its reputation to drop to bad. The difference in payoffs occurs when it has not met with another good requester yet and hence, it does not have the opportunity to restore its reputation or it has met with another good requester, provided the service but remains a bad reputation according to the punishment probability $\alpha$. This makes it lose the instant payoff $b$ at that time period as the requester because of the bad reputation. The utility loss $X$ in the current period can be recursively calculated by

$$X=(1-\rho)\beta X+\rho\left(b+(1-\alpha)\beta X\right) \tag{7}$$

Hence,

$$X=\frac{\rho b}{1-\beta(1-\rho\alpha)} \tag{8}$$

To make following the social strategy incentive compatible, it should be larger than the cost $c$,

$$\frac{\beta\rho b}{1-\beta(1-\rho\alpha)}\geq c \tag{9}$$

which then yields the condition on the user's belief

$$\rho\geq\frac{1-\beta}{\beta(\gamma-\alpha)}\triangleq\rho_G \tag{10}$$

Hence, the provider with a good reputation follows the social strategy only if $\rho\geq\rho_G$

(2) Next we consider the incentives of the bad users. If the requester has a bad reputation, it is also obvious that the provider will follow the social strategy and play defect. If the requester has a good reputation, the decision depends on the provider's belief of the social reputation. At the decision point, the expected payoff is $\pi(\sigma)=-c$ if the provider follows the social strategy and is $\pi(\sigma')=0$ if it deviates. Following the social strategy increases its reputation to 1 with probability $\alpha$. If the realization is that the typical user remains bad, the expected future utility is the same as it deviates. Hence, we only need to consider the realization that the typical user restores a good reputation. Then the analysis is similar to the first case. The loss in the future utility needs to satisfy the following to make incentive compatible for users to follow the social strategy,

$$\frac{\alpha\beta\rho b}{1-\beta(1-\rho\alpha)}\geq c \tag{11}$$

which yields,

$$\rho\geq\frac{1-\beta}{\beta\alpha(\gamma-1)}\triangleq\rho_B \tag{12}$$

The provider with a bad reputation follows the social strategy only if $\rho\geq\rho_B$. ■

The above proposition is consistent with our intuitions: the user will only follow the social strategy if it believes the society is in a sufficiently good status. However, it provides more insights about users' behaviors: (1) Because both $\rho_G, \rho_B$ are strictly positive, there are always a positive fraction of users who deviate because of the heterogeneous beliefs; (2) Increasing the punishment (smaller $\alpha$) prevents fewer good users from deviating (as $\rho_G$ becomes smaller) while it gives more bad users incentives to deviate (as $\rho_B$ becomes larger); (3) The incentive for bad users to cooperate is always no larger than that for good users since $\rho_G \le \rho_B$. The following corollary is a direct result if both belief thresholds are larger than 1 and hence, no user cooperates.

**Corollary 1.** No cooperation can be sustained if

$$\frac{1-\beta}{\beta(\gamma-\alpha)} > 1. \tag{13}$$

The above condition highlights that when users are too impatient (small $\beta$), the benefit-to-cost ratio is too small (small $\gamma$) or the punishment is too mild (large $\alpha$), no cooperation can be sustained. However, when designing information exchange systems, we are interested in sustaining cooperation among the self-interested users and hence, we next derive conditions and the associated system designs to achieve this.

*B. Dynamics and Equilibrium*

Suppose initially the social reputation is $\rho_s$. Limited observations induce heterogeneous beliefs $f(\rho \mid \rho_s)$ among users. Users optimize their strategies $a^*$ according to Proposition 2; these strategies induce dynamics in the new social reputation $\Phi(\rho_s, a^*)$. The equilibrium requires a consistency check: the steady state social reputation remains invariant, i.e.

$$\rho_s = \Phi(\rho_s, a^*) \tag{14}$$

**Definition 1.** (Bayesian-Nash equilibrium) Given $\beta, \gamma, \alpha, M$, let $\rho_s$ be a social reputation, $f(\rho \mid \rho_s)$ be the induced belief distribution due to limited observations, and $a^*$ be the strategy for the users. We say that $(\rho_s, f, a^*)$ constitutes an equilibrium if

1. Users adopt the optimal strategy $\alpha^*$ to maximize their expected utilities (as in Proposition 2).
2. The invariant property holds $\rho_s = \Phi(\rho_s, a^*)$.

It is worth noting that the users' optimal strategy does not rely on the current social reputation $\rho_s$ since the threshold beliefs are only functions of $\beta, \gamma, \alpha$ but not $\rho_s$. However, because the belief distribution is induced by $\rho_s$, the fraction of users who follow the social strategy is thus influenced by $\rho_s$, which in turn determines the social reputation in the next period. If we denote $\Delta(\rho_s) = \Phi(\rho_s, a^*) - \rho_s$ as the change in the social reputation, we can calculate it as follows:

$$\Delta(\rho_s) = \underbrace{\alpha(1-\rho_s)\rho_s F(\rho \geq \rho_B \mid \rho_s)}_{\text{bad to good}} - \underbrace{\rho_s^2 F(\rho \leq \rho_G \mid \rho_s)}_{\text{good to bad}}, \tag{15}$$

with

$$F(\rho \geq \rho_B \mid \rho_s) = \int_{\rho=\rho_B}^{1} f(\rho \mid \rho_s) d\rho, \qquad F(\rho \leq \rho_G \mid \rho_s) = \int_{\rho=0}^{\rho_G} f(\rho \mid \rho_s) d\rho \tag{16}$$

The first part in (15) is the fraction of users whose reputations change from bad to good and the second part is the fraction of users whose reputations change from good to bad. To constitute an equilibrium, it is sufficient and necessary that $\Delta(\rho_s) = 0$. However, we are more interested in whether such an equilibrium is stable if there are some disturbances (e.g. small reputation update errors). The following proposition provides the condition for the stable equilibrium.

**Proposition 3**: (stable equilibrium) The equilibrium with $\rho_s$ is stable if and only if

$$\Delta(\rho_s) = 0 \quad \text{and} \quad \frac{d\Delta(\rho_s)}{d\rho_s} < 0 \tag{17}$$

**Proof**: $\Delta(\rho_s) = 0$ is the condition for equilibrium. Because $\Delta(\rho_s)$ is continuous in $\rho_s$, it is also sufficient and necessary that the first derivative is negative. ■

Now we study the conditions under which the stable equilibrium exists.

**Proposition 4**: Given $\beta, \gamma, \alpha, M$, the existence of the stable equilibrium depends on $\rho_B$.

1. If $\rho_B > 1$, $\rho_s = 0$ is the unique stable equilibrium.

2. If $\rho_B \leq 1$, there exists at least one stable equilibrium $\rho_s \in (0,1)$.

**Proof**: (1) If $\rho_B > 1$, $F(\rho \geq \rho_B \mid \rho_s) = 0$ for all $\rho_s$ and hence, $\Delta(\rho_s) \leq 0$. Equality holds only for $\rho_s = 0$. It is also obvious that the first derivative at $\rho_s = 0$ is negative, therefore it is the only stable equilibrium.

(2) If $\rho_B \leq 1$, we see that $\Delta(\rho_s = 1) < 0, \Delta(\rho_s \to 0) > 0$. Because $\Delta(\cdot)$ is a continuous function in $\rho_s$, it is guaranteed that there exists at least one solution to $\Delta(\rho_s) = 0$ and the first derivative is negative. Moreover, notice that $\rho_s = 0$ is not a stable equilibrium. ■

Proposition 4 proves that neither full efficiency nor zero efficiency occurs in the stable equilibrium in the limited observations case. As we will see later, the actual efficiency will depend on the punishment severity which needs to be carefully designed by the system designer. Before proceeding to that, we compare the achievable efficiency for limited observations case with that for the unlimited observations case to illustrate the different design aspects.

*C. Unlimited observations*

In this subsection, we investigate how the system evolves if users make unlimited observations (i.e. $f(\rho \mid \rho_s) = I(\rho - \rho_s)$) to illustrate why the system design should be different than in the limited observations case. Suppose the system starts with an initial social reputation $\rho_s^0 \in [0,1]$, we are interested in which long-run state $\rho_s^{t\to\infty}$ that the system will be trapped in.

**Proposition** 5: With unlimited observations, the long-run system state is

(1) If $\rho_s^0 \geq \rho_B, \rho_s^{t\to\infty} = 1$; (2) If $\rho_s^0 \leq \rho_G, \rho_s^{t\to\infty} = 0$; (3) If $\rho_G < \rho_s^0 < \rho_B, \rho_s^{t\to\infty} = \rho_s^0$.

**Proof**: Omitted due to space limitation. The proof can be found in [27]. ■

We see that, in unlimited observations case, appropriately choosing the initial social reputation can lead to full efficiency while starting from the wrong initial social reputation leads to zero efficiency regardless of the choice of $\alpha$. This is quite different from the limited observations case where the full efficiency can never be achieved while zero efficiency also does not occur in a stable equilibrium. The achievable efficiency depends on the punishment severity of the social norm and hence, this needs to be carefully designed as discussed in the next section.

## IV. Optimal Punishment Design

The minimum social reputation beliefs $\rho_G, \rho_B$ that sustain cooperation are determined by the punishment. The harsher the punishment is (smaller $\alpha$), fewer good providers deviate while also fewer bad providers cooperate to restore their reputations. Hence, when designing the punishment, the tension between the incentives to the good and bad users needs to be considered. In this section, we characterize the impact of punishment on the achievable system efficiency. In this paper we are interested in maximizing the cooperation among the users and hence, we use the social reputation, i.e. the fraction of good users in the system, as the efficiency metric.

The objective of the system designer in our model is to choose the optimal punishment $\alpha$, given the network environment parameters $\beta, b, c, M$ such that the social reputation is maximized (hence the probability that users cooperate is also maximized which leads to the maximized social welfare). Formally, the design problem is to solve

$$\begin{aligned} &\underset{\alpha}{\text{maximize}} \quad \rho_s \\ &\text{subject to} \quad \Delta(\rho_s) = 0, \ \frac{d\Delta(\rho_s)}{\rho_s} < 0 \end{aligned} \tag{18}$$

In the following, we establish bounds on the achievable efficiency.

**Proposition 6**. Fix $\beta, \gamma, M$ and fix $\alpha$, then the robust equilibrium $\rho_s^*$ is bounded as follows

$$\frac{\alpha(1-\rho_B)^{M+1}}{\alpha(1-\rho_B)^{M+1} + \left(1-(1-\rho_G)^{M+1}\right)} \le \rho_s^* \le \frac{\alpha\left(1-\rho_B^{M+1}\right)}{\alpha\left(1-\rho_B^{M+1}\right) + \rho_G^{M+1}} \tag{19}$$

where $\rho_G, \rho_B$ are determined in Proposition 2.

**Proof:** Omitted due to space limitation. The proof can be found in [27]. ■

**Corollary 2**. Fix $\beta, \gamma, M$, for large $\gamma$, the stable equilibrium $\rho_s^*$ is bounded away from 1,

$$\rho_s^* \le 1 - \left(\frac{1-\beta}{\beta(\gamma-1)}\right)^{M+1}, \forall \alpha \in [0,1] \tag{20}$$

**Proof**: Simply combining Proposition 5 and the fact that the upper bound is increasing in $\alpha$ when $\gamma$ is large yields the result. The right hand side is derived by choosing $\alpha = 1$. ■

The above result shows that the upper bound depends on the granularity of observations. If the system designer wants to achieve a higher efficiency, it is necessary that users are able to make more observations to acquire more accurate reputation distribution information. (Though having more observations may not be the sufficient condition.) In some systems, the number of observations can be designed by the designer. For example, the website designer may only allow users to access the reputations of a limited number of other users due to privacy and security concerns. Therefore the tradeoff between efficiency and privacy needs to be carefully considered. However, in this paper, we assume that the number of observations is exogenously determined.

The mildest punishment maximizes the efficiency upper bound for large $\gamma$. In the following we consider several specific cases of limited observations which induce different user belief distributions and show that the mildest punishment does indeed maximize the efficiency.

*A. Example 1: $M = 0$ (constant belief distribution)*

We consider the simplest case $M = 0$, i.e. users have no observation. In the belief model that we use, $M = 0$ corresponds to the case that users have a (constant) uniform belief over all possible social reputations, namely $f(\rho \mid \rho_s) = 1$ and

$$F(\rho \geq \rho_B \mid \rho_s) = 1 - \rho_B, F(\rho \leq \rho_G \mid \rho_s) = \rho_G \tag{21}$$

For this simple case, we are able to explicitly solve the unique stable equilibrium.

$$\rho_s^* = \frac{\alpha(1-\rho_B)}{\alpha(1-\rho_B)+\rho_G} = \frac{\alpha - \frac{1-\beta}{\beta(\gamma-1)}}{\alpha - \frac{1-\beta}{\beta(\gamma-1)} + \frac{1-\beta}{\beta(\gamma-\alpha)}} \tag{22}$$

It is equivalent to consider the maximization problem,

$$\max_{\alpha} \left(\alpha - \frac{1-\beta}{\beta(\gamma-1)}\right)(\gamma-\alpha) \tag{23}$$

The objective function is a quadratic function. The maximum is achieved at

$$\alpha^* = \min\left\{\frac{\gamma\beta(\gamma-1)+1-\beta}{2\beta(\gamma-1)}, 1\right\} \tag{24}$$

If the benefit-to-cost ratio $\gamma$ is large, the first term in (24) is also larger than 1. Such condition can be easily satisfied (e.g. $\gamma > 2$). Hence, in most scenarios, choosing the mildest punishment, namely $\alpha = 1$, is optimal for the efficiency maximization.

**Proposition 7**: Fix $\beta, \gamma$ and for $M = 0$, the optimal punishment rule is

$$\alpha^* = \min\left\{\frac{\gamma\beta(\gamma-1)+1-\beta}{2\beta(\gamma-1)}, 1\right\} \tag{25}$$

and the induced stable equilibrium is

$$\rho_s^* = \begin{cases} \dfrac{\frac{1}{4}\left(\gamma - \frac{1-\beta}{\beta(\gamma-1)}\right)^2}{\frac{1}{4}\left(\gamma - \frac{1-\beta}{\beta(\gamma-1)}\right)^2 + \frac{1-\beta}{\beta}}, & \text{if } \dfrac{\gamma\beta(\gamma-1)+1-\beta}{2\beta(\gamma-1)} < 1 \\ 1 - \dfrac{1-\beta}{\beta(\gamma-1)}, & \text{if } \dfrac{\gamma\beta(\gamma-1)+1-\beta}{2\beta(\gamma-1)} \geq 1 \end{cases} \tag{26}$$

**Proof**: Simply solving (25) for $\alpha \in [0,1]$ yields the result. ■

*B. Example 2: $M = 1$ (linear belief distribution)*

In this subsection we consider the case $M = 1$. For example, users observe the reputation of one other user by sampling the system. It can also be interpreted as that users have linear belief distribution regarding the true social reputation. The belief function thus is given by

$$f(\rho \mid \rho_s) = \rho_s f_{1,1}(\rho) + (1-\rho_s) f_{0,1}(\rho) = 2\left(1-\rho_s + (2\rho_s - 1)\rho\right) \tag{27}$$

Note $f_{1,1}(\rho) = \rho,\ f_{0,1}(\rho) = 1-\rho$. Hence, the cumulative belief functions are linear in $\rho_s$,

$$F(\rho \geq \rho_B \mid \rho_s) = (1-\rho_B)^2 + 2\rho_B(1-\rho_B)\rho_s \tag{28}$$

$$F(\rho \leq \rho_G \mid \rho_s) = \rho_G(1-\rho_G) - 2\rho_G(1-\rho_G)\rho_s \tag{29}$$

To solve $\Delta(\rho_s) = 0$, it is equivalent to solve $\Delta(\rho_s)/\rho_s = 0$ for $\rho_s \neq 0$. Let $g(\rho_s) = \Delta(\rho_s)/\rho_s$.

$$g(\rho_s) = \alpha\left((1-\rho_B)^2 + 2\rho_B(1-\rho_B)\rho_s\right)(1-\rho_s) - \left(\rho_G(2-\rho_G) - 2\rho_G(1-\rho_G)\rho_s\right)\rho_s \tag{30}$$

The above function is a quadratic function regarding $\rho_s$. It is difficult to solve the stable equilibrium and even more difficult to analyze the impact of punishment directly. In the following, we instead first establish tighter upper and lower bounds of the efficiency in the stable equilibrium than the general bounds given by (26) when $\gamma$ is large. Using the new bounds we are able to derive the optimal punishment based on which optimal social norms can be designed.

**Proposition 8**. Fix $\beta, \gamma$ and fix $\alpha$, $M = 1$, for large $\gamma$, the stable equilibrium $\rho_s^*$ is bounded by

$$\frac{\alpha(1-\rho_B)^2}{\alpha(1-\rho_B)(1-3\rho_B) + \rho_G(2-\rho_G)} \leq \rho_s^* \leq \frac{\alpha(1-\rho_B)^2}{\alpha(1-\rho_B)^2 + \rho_G^2} \tag{31}$$

**Proof**: For large $\gamma$, the belief thresholds $\rho_G$ and $\rho_B$ are approximated by

$$\rho_G = \frac{1-\beta}{\beta\gamma}, \qquad \rho_B = \frac{1-\beta}{\alpha\beta\gamma} \tag{32}$$

(1) We first establish the upper bound. Note that the quadratic coefficient of (30) is

$$2\left(\rho_G(1-\rho_G) - \alpha\rho_B(1-\rho_B)\right) = 2\left(\frac{1-\beta}{\beta\gamma}\right)^2\left(\frac{1}{\alpha} - 1\right) > 0 \tag{33}$$

Hence, $g(\rho_s)$ is a convex quadratic function. Because there must be one and only one root that lies in $(0,1)$, it is upper bounded by

$$\rho_s^* \leq \frac{g(0)}{g(0) - g(1)} = \frac{\alpha(1-\rho_B)^2}{\alpha(1-\rho_B)^2 + \rho_G^2} \tag{34}$$

Because $(1-\rho_B)^2 < 1-\rho_B^2$, this upper bound is tighter than the general upper bound.

(2) Next we establish the lower bound. Note the slope at $\rho_s = 0$ of $g(\rho_s)$ is

$$\alpha\left(1-\rho_B\right)\left(3\rho_B-1\right)-\rho_G\left(2-\rho_G\right) \quad (35)$$

By the convexity, the root in $(0,1)$ is lower bounded by

$$\rho_s^* \geq \frac{g(0)}{\alpha\left(1-\rho_B\right)\left(1-3\rho_B\right)+\rho_G\left(2-\rho_G\right)}=\frac{\alpha\left(1-\rho_B\right)^2}{\alpha\left(1-\rho_B\right)\left(1-3\rho_B\right)+\rho_G\left(2-\rho_G\right)} \quad (36)$$

Because $\left(1-3\rho_B\right)<\left(1-\rho_B\right)$, the lower bound is tighter than the general bound. ■

The upper bound in the above proposition in fact has more implications for the optimal punishment design. For large $\gamma$, in order to maximize the upper bound, it is equivalent to consider the following maximization problem

$$\max_{\alpha} \alpha\left(1-\rho_B\right)^2 \quad \text{or} \quad \max_{\alpha} \alpha\left(1-\frac{1-\beta}{\alpha\beta\gamma}\right)^2 \quad (37)$$

Expanding the objective function in (37), we get

$$\alpha\left(1-\frac{1-\beta}{\alpha\beta\gamma}\right)^2=\alpha+\frac{1}{\alpha}\left(\frac{1-\beta}{\beta\gamma}\right)^2-2\frac{1-\beta}{\beta\gamma}. \quad (38)$$

Remember that we need to ensure that $\rho_B<1$ since otherwise the only robust equilibrium is 0 according to Proposition 4 and hence, the feasible $\alpha$ needs to satisfy

$$\alpha \geq \frac{1-\beta}{\beta\gamma}. \quad (39)$$

Choosing $\alpha=1$ maximizes (38) and hence, it maximizes the upper bound for all feasible $\alpha$. Note that for $\alpha=1$, the upper bound is indeed the actual efficiency because the upper and lower bounds are identical. Therefore, $\alpha=1$ maximizes the efficiency of the stable equilibrium. The following proposition restates this result and also determines the social reputation in equilibrium.

**Proposition 9**. Fix $\beta,\gamma,M=1$ for large $\gamma$, the stable equilibrium $\rho_s^*$ is maximized by choosing $\alpha^*=1$, and the optimal solution is

$$\rho_s^*=\frac{\left(1-\frac{1-\beta}{\beta\gamma}\right)^2}{\left(1-\frac{1-\beta}{\beta\gamma}\right)^2+\left(\frac{1-\beta}{\beta\gamma}\right)^2} \quad (40)$$

Note that this stable equilibrium efficiency is close to 1 when $\gamma$ is large or $\beta$ is close to 1. For the higher order cases, i.e. values of $M$ other than 0 and 1, it is rather difficult to derive any analytical results. We will investigate this in the simulation section numerically. However, from the analysis for the two specific examples, some design insights can still be drawn: when the benefit-to-cost ratio is large (1), it is optimal to choose $\alpha = 1$, which is the mildest punishment possible, and (2) a larger $M$ leads to a higher social reputation for this choice of punishment.

## V. SIMULATIONS

In this section, we provide some simulation results. Fig. 3 illustrates the system evolution for various environments. It is shown that the system quickly converges to the stable state. In this set of simulations $M = 0,1,2$, the stable equilibrium is unique and hence, any initial state converges to the same stable equilibrium. However, there can also be multiple stable equilibria in which case different initial states converge to different stable equilibria. As we show in Fig. 4 for $M = 6$, there are two stable equilibria. If the system starts with a high initial state, it eventually has a high social reputation while if the initial social reputation is low, the final social reputation is also low. Fig. 5 shows the case with perfect information, i.e. $M = \infty$. If the initial social reputation is higher than $\rho_B$, no matter which $\alpha$ the system designer chooses, the system achieves full efficiency, i.e. all agents have good reputations; if the initial social reputation is lower than $\rho_G$, the system achieves zero efficiency, i.e. all agents have bad reputations; for the initial social reputation that lie between $[\rho_G, \rho_B]$, the system stays in the same state.

Fig. 6 illustrates the impact of $M$. As we see, more observations do not necessarily lead to better performance for a given punishment. In fact, the bounds established in (19) does not tell the monotonicity regarding $M$. However, for a larger $\alpha$, more observations do have a better performance. Because $\alpha = 1$ is often the optimal choice, basically $M$ should be larger to achieve a higher efficiency. In order to obtain the better performance, users need to have more accurate information of the reputation distribution. Fig. 7 further compares the simulated optimal equilibria with the bounds established by (20). The established bounds are close to the simulation points and the performance becomes quite close to full efficiency as $M$ increases.

Fig. 8 and Fig. 9 illustrate the impact of the benefit-to-cost ratio $\gamma$ and the discount factor $\beta$. For a given punishment probability, the system performance improves with $\gamma$. Moreover, for all simulated values of $\gamma$, choosing $\alpha = 1$ generates the best performance. The discount factor $\beta$ has a similar impact as $\gamma$: Larger $\beta$ leads to a better performance and choosing $\alpha = 1$ generates the highest efficiency. Even though the fact that the mildest punishment is optimal may seem counter-intuitive, this finding can be easily explained as follows. Punishment is often used to prevent users from misbehaving. When users are good, harsher punishments impose greater threat on these users if they would deviate. Hence, it may seem that harsher punishments are needed to obtain a better performance. However, this intuition is only valid when all users are on the equilibrium path, i.e. they always follow the social strategy. For the limited observations scenario, there are always a positive fraction of users who deviate no matter what the punishment is. Once users are in the punishment phase, harsher punishment becomes a disincentive for them to restore their reputations. As we show that punishment has much greater impact on the belief threshold for bad users, the system eventually will be in an equilibrium with a lower efficiency.

## VI. Conclusions

In this paper, we design the optimal social norm protocol for information exchange systems where users have heterogeneous beliefs due to limited observations of the system. First, a Bayesian belief model is proposed to model the belief heterogeneity. Second, the optimal provider strategy is shown to have a threshold property: users cooperate only when they have sufficient "trust" in the system (i.e. believe that sufficient users are cooperating). Finally, the impact of the punishment severity on the stable equilibrium and the achievable system efficiency is rigorously studied. When users can make unlimited observations, full or zero efficiency occurs in the stable equilibrium. However, in the more realistic limited observations scenario, full efficiency can never be achieved and different punishment strategies lead to different stable equilibria having different efficiencies. We show that choosing the mildest punishment is optimal for many interesting scenarios and support this finding with both analytical and simulation results.

## References

[1] O. Loginova, H. Lu, and X. H. Wang, "Incentive schemes in peer-to-peer networks," *The B.E. Journal of Theoretical Economics*, Oct. 2008.

[2] K. Ranganathan, M. Ripeanu, A. Sarin, and I. Foster, "Incentive mechanism for large collaborative resource sharing," In *Proceedings of IEEE International Symposium on Cluster Computing and the Grid*, 2004.

[3] P. Antoniadis, C. Courcoubetis, and B. Strulo, “Comparing economics incentives in peer-to-peer networks,” *Computer Networks*, 46(1):133-146, 2004.

[4] E. J. Friedman, J. Y. Halpern, and I. A. Kash, “Efficiency and nash equilibria in a scrip system for P2P networks,” In *Proceedings of the Seventh ACM Conference on Electronic Commerce*, 2006.

[5] I. A. Kash, E. J. Friedman, and J. Y. Halpern, “Optimizing scrip systems: Efficiency, crashes, hoarders, and altruists,” In *Proceedings of the Eighth ACM Conference on Electronic Commerce*, 2007.

[6] R. Landa, D. Griffin, R. Clegg, E. Mykoniati, and M. Rio, “A sybilproof indirect reciprocity mechanism for peer-to-peer networks,” In *Proceedings of INFOCOM*, 2009.

[7] J. Xu, M. van der Schaar, and W. Zame, “Designing exchange for online communities,” *Technical report* available at http://arxiv.org/abs/1108.5871.

[8] R. L. Trivers, “The evolution of reciprocal altruism,” *Quarterly review of biology*, 46(1):35-57, 1971.

[9] R. D. Alexander, *The Biology of Moral Systems*, Aldine de Gruyter, New York, 1987.

[10] H. Masum and Y. Zhang, “Manifesto for the reputation society,” *First Monday*, 9(7), 2004.

[11] S. Kamvar, M. T. Schlosser, and H. G. Molina, “The eigentrust algorithm for reputation management in P2P networks,” In *Proceedings of 12th International Conf. on World Wide Web*, 2003.

[12] A. Ravoaja and E. Anceaume, “Storm: a secure overlay for P2P reputation management,” In *Proceedings of 1st International Conf. on Self-Adaptive and Self-Organizing Systems*, 2007.

[13] S. Ba and P. Pavlou, “Evidence of the effect of trust building technology in electronic markets: price premiums and buyer behavior,” *MIS Quart*, 26(3):243-268, 2002.

[14] P. Resnick, R. Zeckhauser, “Trust among strangers in internet transactions: empirical analysis of eBay’s reputation system,” *Advances in Applied Microeconomics*, 11, 2002.

[15] P. Resnick, R. Zeckhauser, J. Swanson, and K. Lockwood, “The value of reputation on eBay: a controlled experiment,” *Exp Econ*, 9:79-101, 2006.

[16] C. Keser, “Experimental games for the design of reputation management systems,” *IBM Systems Journal*, 42(3):498-506, 2003.

[17] C. Dellarocas, “Reputation mechanism design in online trading environments with pure moral hazard,” Information Systems Research, 16(2):209-230, 2005.

[18] C. Dellarocas, “How often should reputation mechanisms update a trader’s reputation profile?” *Information Systems Research*, 17(3):271-285, 2006.

[19] M. Fan, Y. Tan, and A. B. Whinston, “Evaluation and design of online cooperative feedback mechanism for reputation management,” *IEEE Transactions on Knowledge Data Engineering*, 17(3):244-254, 2005.

[20] G. Zacharia, A. Moukas, and P. Maes, “Collaborative reputation mechanisms in electronic marketplaces,” *Decision Support Systems*, 29(4):371-388, 2000.

[21] Y. Zhang, J. Park, and M. van der Schaar, “Social norms for networked communities,” *Technical report* available at http://arxiv.org/abs/1101.0272, 2011.

[22] G. Ellison, “Cooperation in the prisoner’s dilemma with anonymous random matching,” *Review of Economic Studies*, 61(3):567-588, 1994.

[23] M. Kandori, “Social norms and community enforcement,” *Review of Economic Studies*, 59(1):63-80, 1992.

[24] M. Okuno-Fujiwara and A. Postlewaite. “Social norms and random matching games,” *Games and Economic Behaviors*, 9(1):79-109, 1995.

[25] K. Iyer, R. Johari, M. Sundararajan, “Mean field equilibria of dynamic auctions with learning,” In *Proceedings of the 12th ACM conference on Electronic Commerce*, 2011.

[26] Y. Zhang and M. van der Schaar, “Influencing the long-term evolution of online communities using social nroms,” *Allerton Conference*, 2011.

[27] Omitted proofs available at http://ee.ucla.edu/~jiexu/documents/appendix_jsac_111215.pdf.

| | Provider | |
|---|---|---|
| | Provide service | Not provide service |
| **Requester** | *b, -c* | *0,0* |

Figure 1. The utility matrix of the gift-giving game.

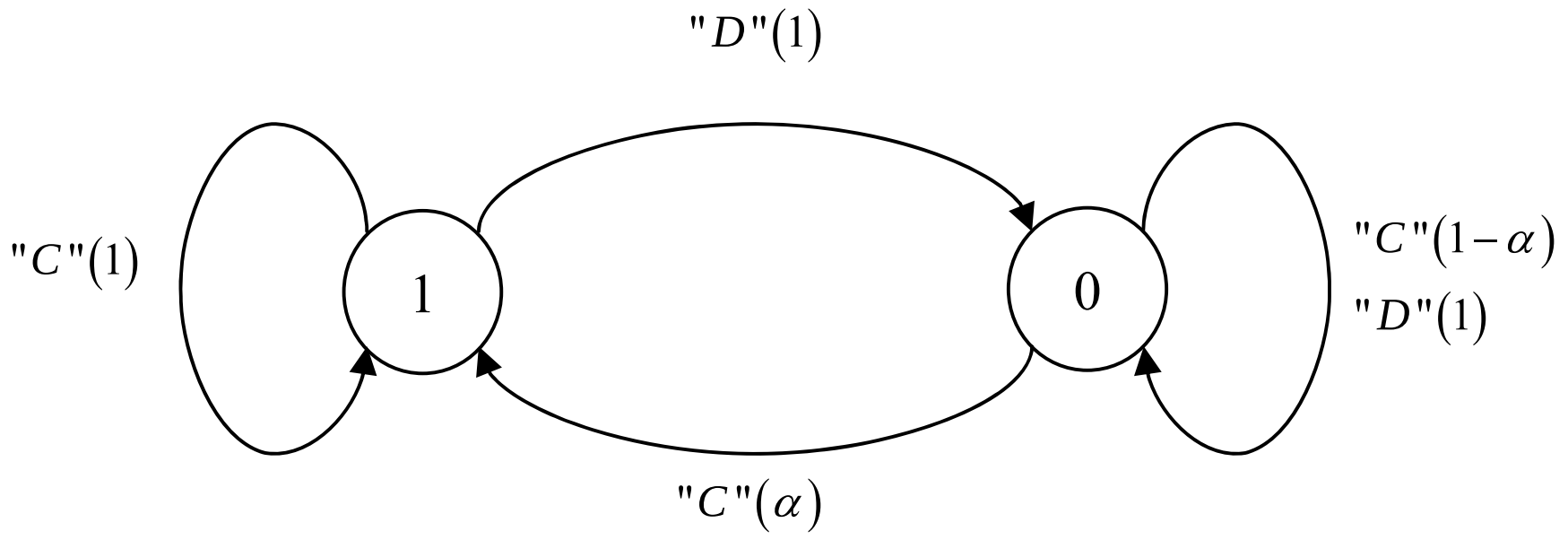

Figure 2. Reputation updating rule.

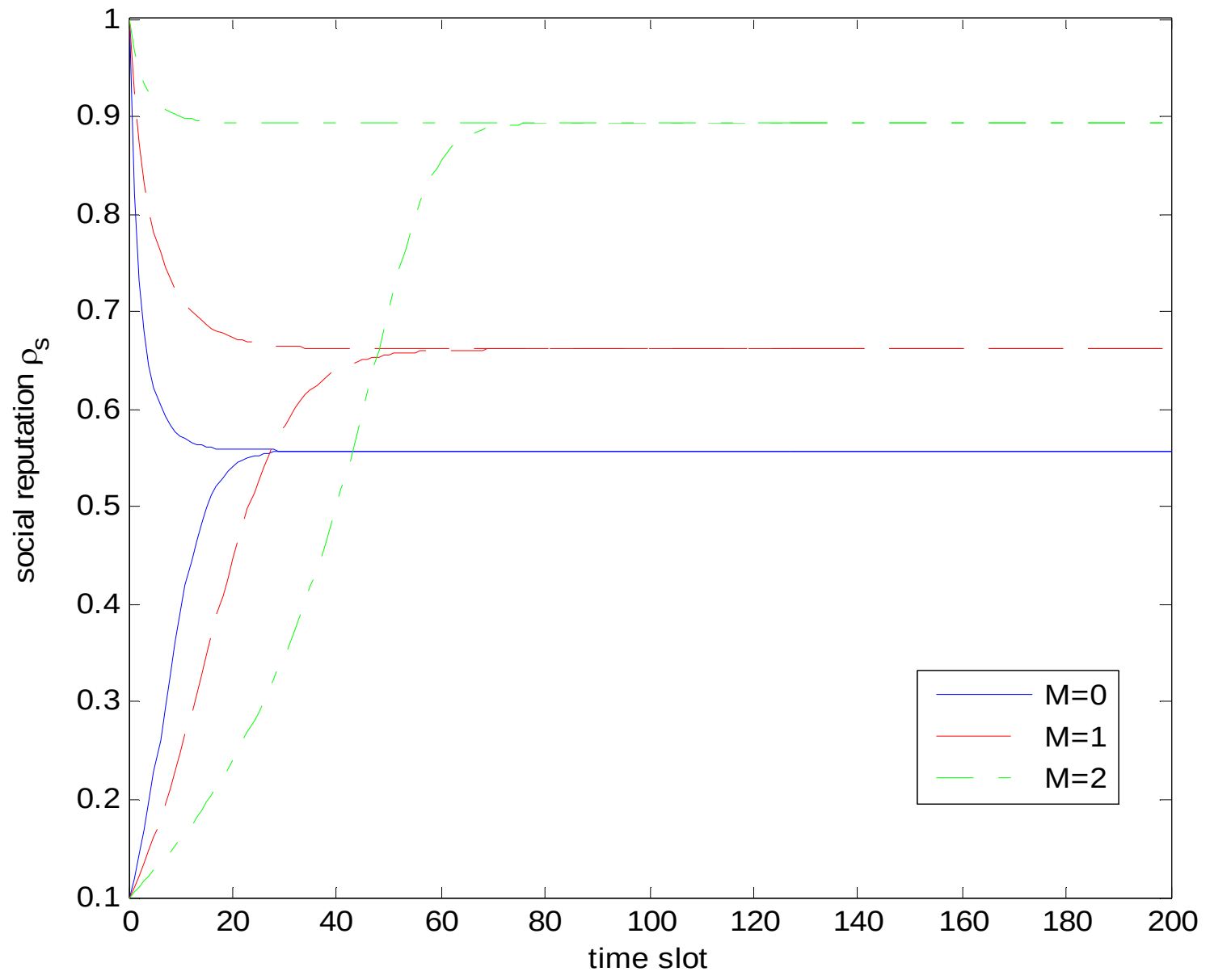

Figure 3. Stable state of the system for $\beta = 0.25, \gamma = 10$ and $\alpha = 0.9$.

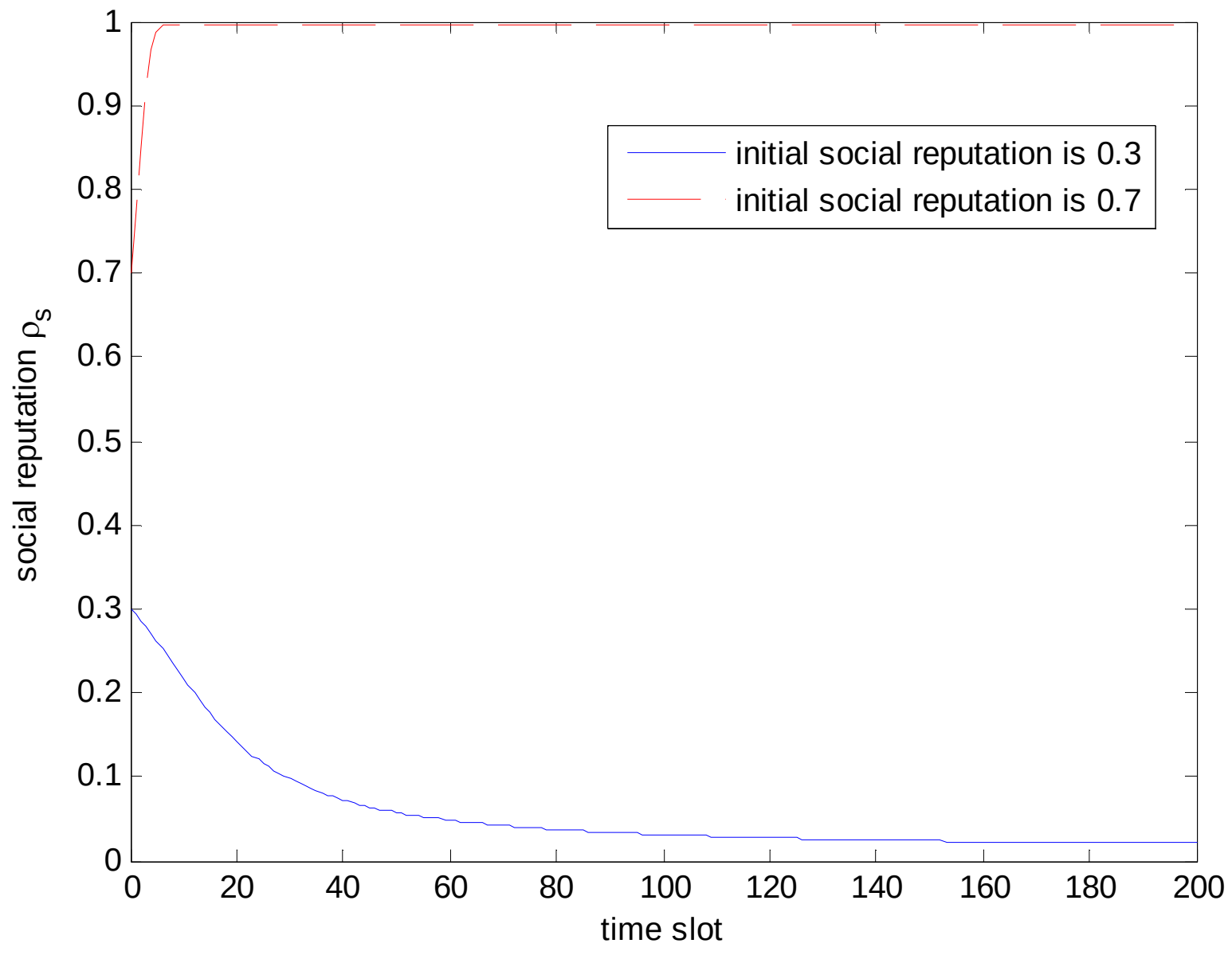

Figure 4. Stable state of the system for $\beta = 0.25, \gamma = 8, M = 6$ and $\alpha = 0.9$.

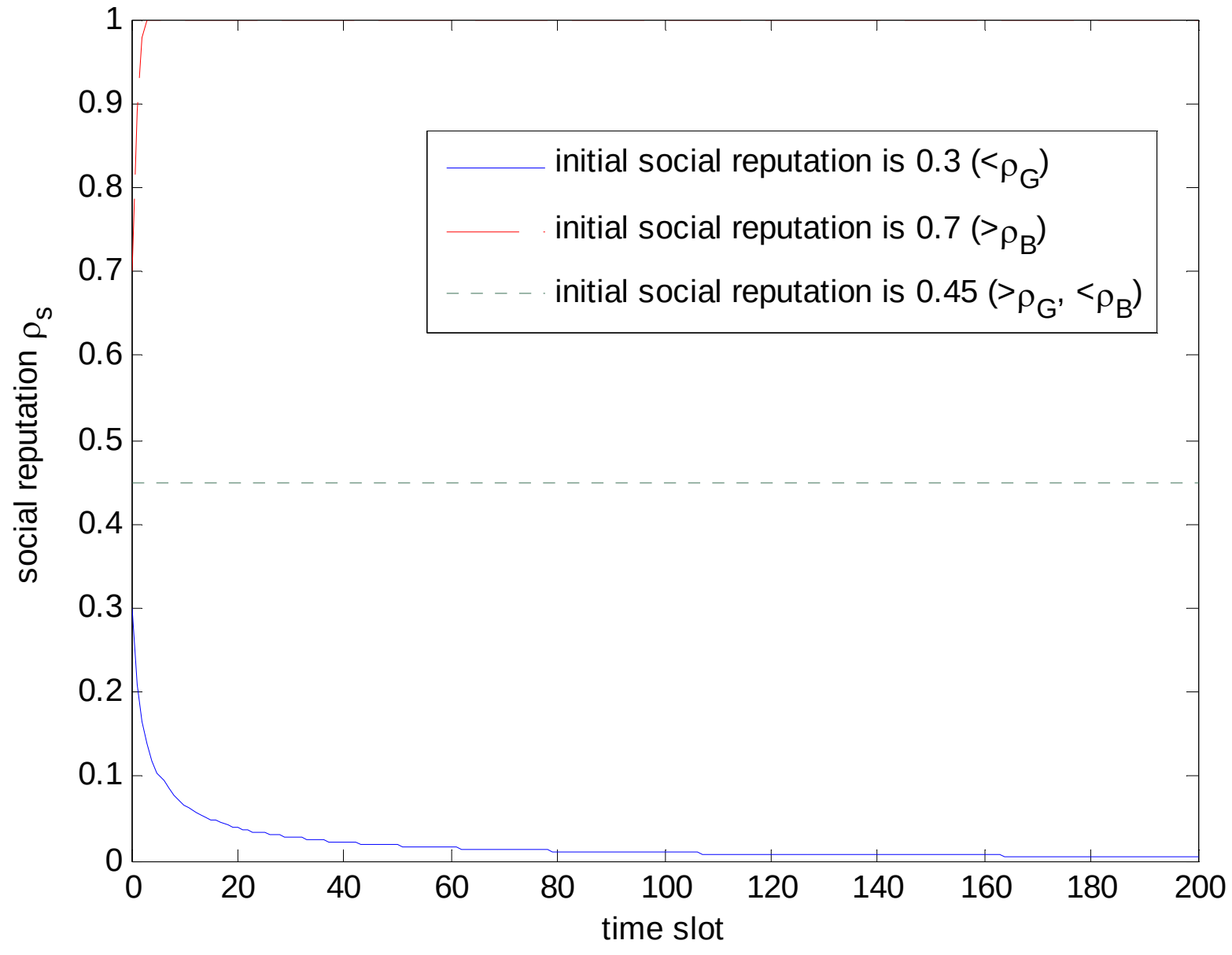

Figure 5. Stable state of the system for $\beta = 0.25, \gamma = 8, M = \infty$ and $\alpha = 0.9$.

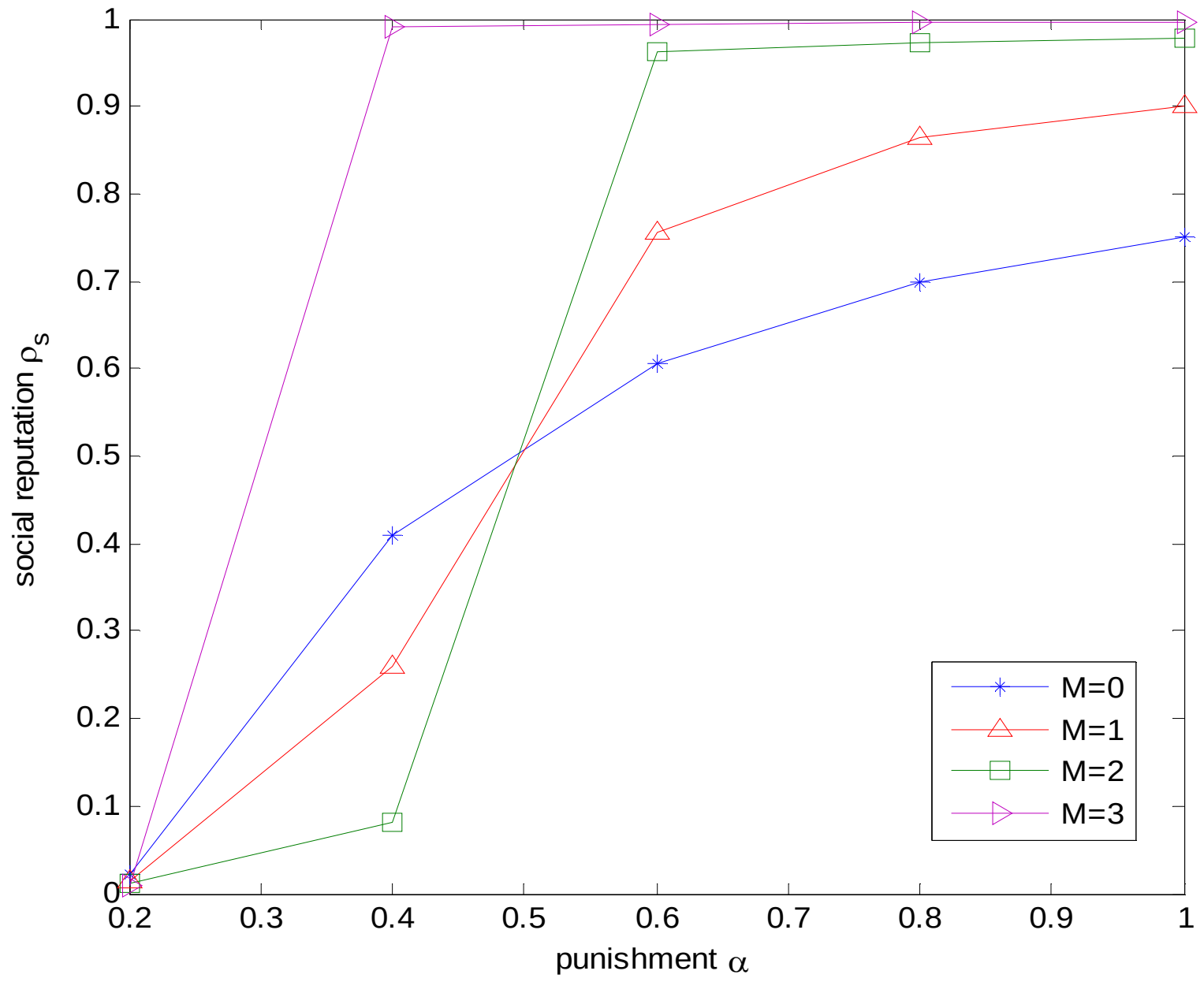

Figure 6. Impact of the observation granularity $M$ (fix $\beta = 0.5, \gamma = 5$).

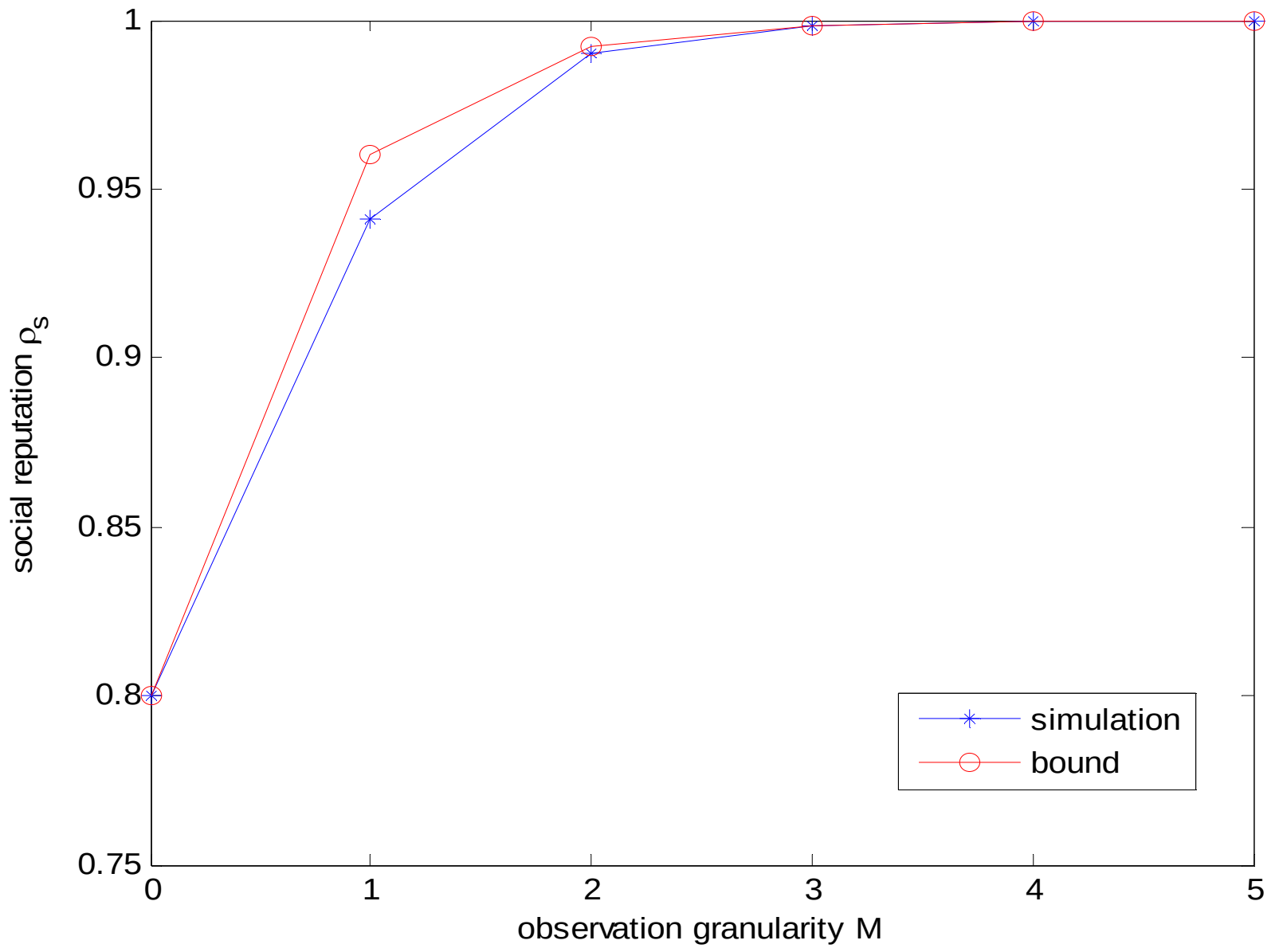

Figure 7. Bounds on the optimal social reputation for various observation granularities. ($\gamma = 6, \beta = 0.5$)

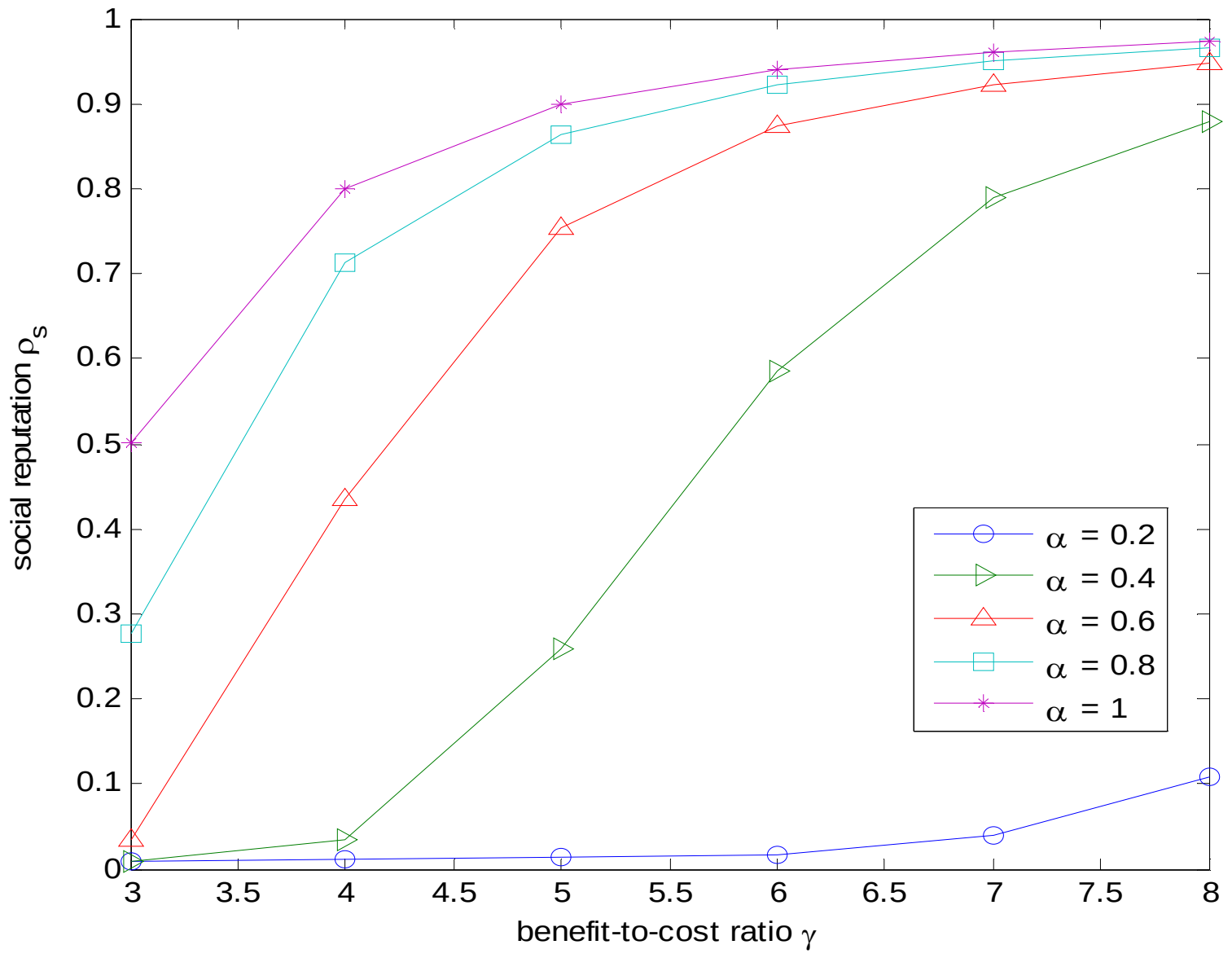

Figure 8. Impact of the benefit-to-cost ratio $\gamma$ (fix $\beta = 0.5, M = 1$).

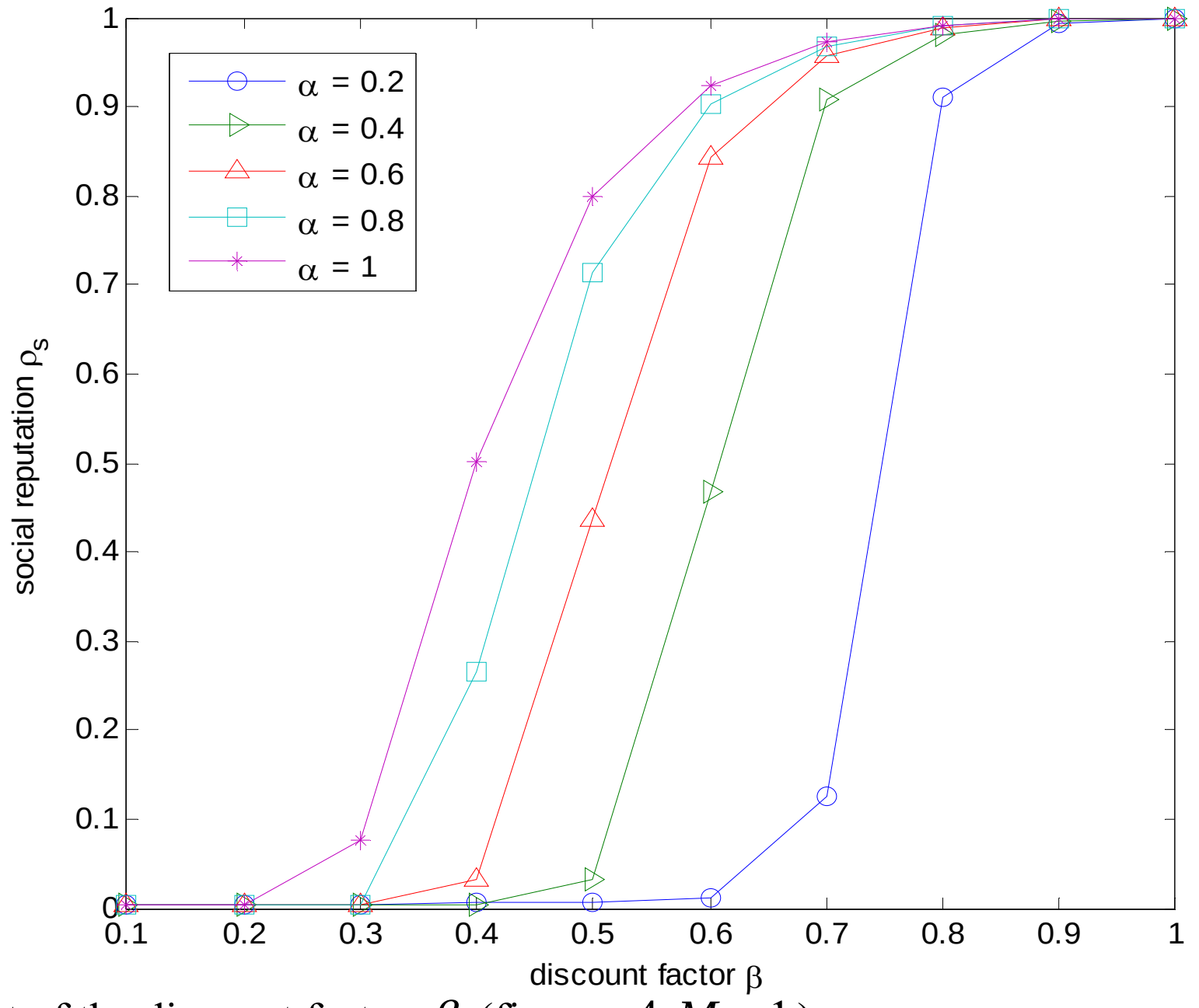

Figure 9. Impact of the discount factor $\beta$ (fix $\gamma = 4, M = 1$).